\documentclass[a4paper, 11pt]{article}

\usepackage{a4wide}
\usepackage{amsmath}
\usepackage{commath}
\usepackage{latexsym}
\usepackage{graphicx}
\usepackage{color}
\usepackage[pdfstartview=FitH]{hyperref}
\usepackage{verbatim}

\title{A stochastic differential equation approach to the analysis of the UK 2016 EU referendum polls}

\author{Trevor Fenner, Mark Levene, and George Loizou \\
Department of Computer Science and Information Systems \\
Birkbeck, University of London \\
London WC1E 7HX, U.K. \\ \{mark,trevor,george\}@dcs.bbk.ac.uk}

\date{} {

\begin{document}

\maketitle

\begin{abstract}

Human dynamics and sociophysics suggest statistical models that may explain and provide us with better insight into social phenomena.
Here we propose a generative model based on a stochastic differential equation that allows us to analyse the polls leading up to the UK 2016 EU referendum. After a preliminary analysis of the time series of poll results, we provide empirical evidence that the beta distribution, which is a natural choice when modelling proportions, fits the marginal distribution of this time series. We also provide evidence of the predictive power of the proposed model.

\end{abstract}


\section{Introduction}

Recent interest in complex social systems, such as social networks, the world-wide-web, messaging networks and mobile phone networks \cite{BARA16}, has led researchers to investigate the processes that could explain the dynamics of human behaviour within these networks. Human dynamics is not limited to the study of behaviour in communication networks, and has a broader remit similar to the aims of {\em sociophysics} \cite{GALA08,SEN14} (also known as {\em social physics}), which uses concepts and methods from statistical physics to investigate social phenomena, opinion formation and political behaviour. A central idea here is that, in the context of statistical physics, individual humans can be thought of as ``social atoms'', each exhibiting simple individual behaviour and possessing very limited intelligence, but nevertheless collectively yielding complex social patterns \cite{BENT11a}.

Social physics has a long history going back to the polymath Quetelet in the 19th century, who applied statistical laws to the study of human characteristics; for example, in deriving the {\em body mass index}, he discovered that body weight is approximately proportional to the square of the body height \cite{EKNO08}. The foundations of 20th century social physics can be attributed to Stewart \cite{STEW50}, whose research was linked to applying gravitational potential theory to the geographic distribution of populations.

\smallskip

Polls impart important information to the public in the lead-up to an election or a referendum, and provide an important ingredient of forecasting methods. However, assessing their accuracy is of major concern due to various sources of variability \cite{CONV86}. Sampling error can typically be quantified by providing confidence intervals \cite{FRAN07}, although it is not the only source of error.
Polls in a given election cycle can be naturally viewed as a time series, and thus be expected to follow a stochastic process, such as
an AR(1) model \cite{CHAT96}. In \cite{WLEZ02} the authors concluded that such a time series model is often not feasible for two reasons. First, the presence of sampling error makes it difficult to obtain reliable parameters for the time series model, and, second, there is generally a lack of sufficient time series data for a given election to enable us to build a robust model. However, in \cite{WLEZ17} it was mentioned that, given a sufficient number of poll results, these could be readily treated as a statistical time series. In the case of the UK EU referendum, also known as the ``Brexit'' referendum, we have a collection of 168 polls, conducted regularly by different pollsters over a period of 10 months leading up to the referendum. We believe that this justifies a fresh look at the time series approach, as presented here, which goes beyond the model suggested in \cite{WLEZ02}. We note that in \cite{WLEZ02,WLEZ17} a novel method was presented to analyse a multitude of polls over the election cycle, across several different elections. One result of this analysis showed convincingly, as one might expect, that polls are generally more accurate the closer they are to the actual election.

\smallskip

We note that a time series model, which captures statistical patterns, is intended to help us gain a better understanding of the data, as we do not have full knowledge of the variables that affect voters' choices. Thus it is meant to complement rather than replace multivariate analysis \cite{HAIR14}, such as the aggregate-level analysis carried out in \cite{GOOD16a} in order to investigate the socio-demographic predictors of the referendum vote.

\smallskip

Another rich source of data nowadays comes from social media such as Twitter data, which is indeed plentiful. Making use of sentiment analysis technology \cite{LIU15}, it was demonstrated in \cite{OCON10} that sentiment correlates highly with polling data.
In \cite{ANUT17}, it was found that opinions based on Twitter were more biased than those gleaned from the polls, when compared with the actual outcome. However, if the biases in social data can be detected, it is possible that the accuracy of election predictions could be improved \cite{BOHA17}.

\medskip

In the context of human dynamics, we have been particularly interested in formulating {\em generative models} in the form of stochastic processes by which complex systems evolve and give rise to power laws or other distributions \cite{FENN15}. This type of research builds on the early work of Simon \cite{SIMO55}, and the more recent work of Barab\'asi's group  \cite{ALBE01} and other researchers.  In recent work \cite{FENN16b,FENN17a}, we have employed a multiplicative model that is designed to capture the essential dynamics of survival analysis applications \cite{KLEI12}. The resulting rank-ordering distribution \cite{SORN96}, the {\em beta-like distribution} (cf. \cite{MART09}), is a discrete analogue of the {\em beta distribution} \cite{GUPT04}.
The beta-like distribution was deployed in \cite{FENN16b} to model constituency-based general election results, while in \cite{FENN17a}
it was utilised to model the regional results in the UK 2016 EU referendum.

\smallskip

Generative models, arising from {\em agent-based modelling} \cite{CONT14}, have played an important role in the sociophysics literature in the context of opinion dynamics \cite{CAST09,SIRB17}. In particular, the voter model and its extensions \cite{CAST09,SIRB17} have applications in explaining and understanding voting behaviour during elections.
A voter model can be described, in its simplest form, as a stochastic process, whereby at each time step an agent decides whether to
hold onto or change its opinion, depending on the opinions of its neighbours. An agent-based herding model of voting behaviour, recently presented in \cite{KONO17}, that models the share of votes across polling stations was shown to follow a beta distribution, in a similar way to the model we present here.

\medskip

Here we direct our attention to modelling the polls leading up to the UK 2016 EU referendum as a time series,
as mentioned earlier. In particular, we make use of {\em stochastic differential equations} (SDEs) \cite{MACK11,EVAN13},
a model widely used in physics and mathematical finance, which can be viewed as a continuous approximation to a discrete process modelling how the polls vary over time. Such a discrete model, using difference equations, has been extensively studied in the context of obtaining numerical solutions to SDEs \cite{IACU08,SAUE13}. Here we are interested in ``mean reverting'' SDEs \cite{HIRS14} for which the time series they describe have stationary solutions with well-known distributions that depend on the form of the underlying SDE \cite{COBB81,BIBB05}.
In particular, we found that the beta distribution \cite{GUPT04} is a good fit to the marginal distribution of the polls time series. This distribution is well-suited to our application for the following reasons: first, the beta distribution is a flexible distribution designed to deal with proportions due to its bounded support (cf. \cite{GUOL14}) and, second, it is the conjugate prior of the binomial distribution and thereby allows us to adjust our beliefs about the true proportions by taking into account the latest opinion poll results.

\smallskip

The main contribution of the paper is to demonstrate empirically that a time series model based on SDEs, with a marginal beta distribution, is suitable for modelling how poll results change over time. Moreover, since models using SDEs can also be used for prediction \cite{JUHL16}, we also consider the predictive power of our model.

\medskip

The rest of the paper is organised as follows.
In Section~\ref{sec:polls} we provide a preliminary analysis of the referendum poll results using the normal confidence interval methodology. In Section~\ref{sec:model} we propose a random walk model for analysing the polling data based on a ``mean reverting'' stochastic differential equation. In Section~\ref{sec:uk} we apply the model to the polls leading up to the UK 2016 EU referendum.
Finally, in Section~\ref{sec:conc} we give our concluding remarks.

\section{Preliminary analysis of the time series of poll results}
\label{sec:polls}

The analysis was done on the results of 168 opinion polls, which were conducted prior to the referendum that took place on 23rd June 2016. Out of the 168 polls, 155 of them also recorded how many people were undecided at the time. The data set was obtained online from \cite{WHAT16}, the first poll being taken on 1st September 2015 and the last one taken the day before the referendum. The mean, standard deviation (Std) and coefficient of variation (CV, defined as Std/Mean) for the polls is shown in Table~\ref{table:mean}; it can be seen that, according to the polls, the Remain campaign was leading, on average, by approximately 3\% during the polling period. In addition, it can be seen that the CV, approximately 11\% for Remain and 10\% for Leave, is rather high, indicating that, according to the polls, the referendum result was far from certain. It is clear that the standard deviation for Undecided is high relative to its mean, giving rise to the very high CV, which is indicative of the volatility of the Undecided vote.

\begin{table}[ht]
\begin{center}
\begin{tabular}{|l|c|c|c|}\hline
Response  & Mean   & Std     & CV     \\ \hline \hline
Remain    & 44.45\% & 4.99\% & 11.23\% \\ \hline
Leave     & 41.63\% & 4.13\% & 9.92\% \\ \hline
Undecided & 14.97\% & 5.42\% & 36.20\% \\ \hline
\end{tabular}
\end{center}
\caption{\label{table:mean} Mean and standard deviation for the polls.}
\end{table}
\smallskip

As a preliminary step, we test the statistical significance of the difference between Remain and Leave for each of the polls, using a 95\% confidence interval for the difference between two proportions from the same population \cite[Equation~3.4]{SEBE13} (see also \cite{SCOT83} and \cite{FRAN07}), given by
\begin{equation}\label{eq:prop}
\hat{p_1} - \hat{p_2} \pm 1.96 \sqrt{\frac{\hat{p_1} + \hat{p_2} - \left( \hat{p_1} - \hat{p_2} \right)^2}{n}},
\end{equation}
where $\hat{p_1}$ is the Remain proportion, $\hat{p_2}$ is the Leave proportion, and $n$ is the sample size.

\smallskip

Overall, in 70 out of the 168 polls, i.e. 41.67\%, the difference between Remain and Leave was significant. Furthermore, in 56 out of those 70 polls, i.e. 80\% of the statistically significant polls, the proportion for Remain was larger than the proportion for Leave. Interestingly, when looking at all of the 168 polls, in 99 of these the proportion for Remain was larger than that for Leave, which is only 58.93\% compared to the 80\% for Remain in the significant polls.
In the actual referendum 33,551,983 people voted, which was a massive turnout of 72.2\% of the electorate. Out of these,
48.11\% voted Remain and 51.89\% voted Leave, which is a statistically significant result according to the test.
Moreover, the difference between Leave and Remain was 3.78\%, and the 95\% confidence interval for the difference, i.e., [3.75\%, 3.81\%], is very narrow.

\smallskip

We next divided the 168 polls into two equal groups, where the first 84 took place from September 2015 until the 22nd March 2016, and the second 84 took place from the 23rd of March  2016 until the day before the referendum. It transpired that for 41 out of the first group of polls, i.e. 48.81\%, the difference between the Remain and Leave proportions was statistically significant, while for the second group it was significant for only 29 polls, i.e. 34.52\%. Out of the 41 significant polls in the first group, Remain was leading in 36 polls, i.e. 87.80\%, while, out of the 29 significant polls in the second group, Remain was leading in 19 polls, i.e. 65.52\%. However, considering the overall poll results, whether significant or not, Remain was leading in 57 polls in the first group , i.e. 67.86\%, whereas Remain was leading in only 42 polls in the second group, i.e. 50\%. This indicates that, although, according to the polls, the gap between Remain and Leave was closing as the referendum approached, it was nevertheless quite likely that Remain would win the final vote.

\smallskip

We also tested whether the proportion of undecided voters during the polling period was significantly different from zero,
using the 95\% confidence interval for a single proportion \cite[ Equation~2.4]{SEBE13}, known as Wald's confidence interval, given by
\begin{equation}\label{eq:wald}
\hat{p} \pm 1.96 \sqrt{\frac{\hat{p} \left(1 - \hat{p}\right)}{n}},
\end{equation}
where $\hat{p}$ is the Undecided proportion and $n$ is the sample size.

\smallskip

In all of the 155 polls that recorded undecided voters, the proportion of undecided voters was significant.
On average 14.97\% of voters in these 155 polls indicated that their vote was undecided, and this vote could have potentially swayed these polls in either direction.

\medskip

We then computed the {\em mean absolute errors} and the {\em root mean square errors} \cite{CHAI14} for Remain and Leave compared to the final results.
The mean absolute error ({\em MAE})  is given by
\begin{equation}\label{eq:mae}
MAE = \frac{\sum_{i=1}^{n} | p_i - f |}{n},
\end{equation}
where $p_i$ is the Remain or Leave proportion in the $i$th poll, $f$ is the Remain or Leave proportion of votes in the actual referendum, and $n$ is the number of polls. The root mean square error ({\em RMSE}) is given by
\begin{equation}\label{eq:rmse}
RMSE = \sqrt{\frac{\sum_{i=1}^{n} \left(p_i - f\right)^2}{n}}.
\end{equation}
\smallskip

The results are shown in Table~\ref{table:err}, where it can be seen that the errors for Leave are approximately twice as large as those for Remain. This is not surprising given the final, somewhat unexpected, result.

\begin{table}[ht]
\begin{center}
\begin{tabular}{|l|c|c|}\hline
Response & {\em MAE}& {\em RMSE}  \\ \hline \hline
Remain   &  5.37\%  &  6.11\% \\ \hline
Leave    & 10.40\%  & 11.16\% \\ \hline
\end{tabular}
\end{center}
\caption{\label{table:err} {\em MAE} and {\em RMSE} for the polls.}
\end{table}
\smallskip

\begin{figure}[ht]
\begin{minipage}{0.5\textwidth}
\includegraphics[scale=0.39]{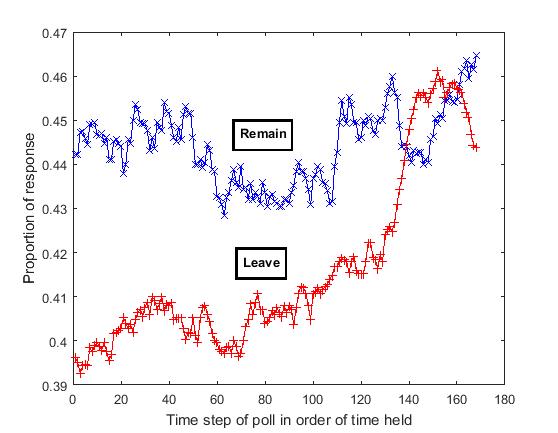}
\end{minipage}
\begin{minipage}{0.5\textwidth}
\includegraphics[scale=0.39]{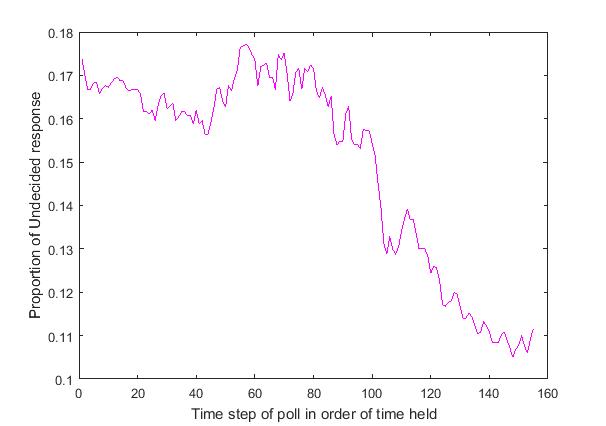}
\end{minipage}
\caption{\label{figure:ma_ts} Moving average with a centred sliding window of 25 time steps for the Remain and Leave (left), and Undecided (right) poll results.}
\end{figure}

When analysing the data, it is also interesting to inspect the moving average \cite{CHAT96} of the polls, as shown in Figure~\ref{figure:ma_ts}, in order to see any trend. In this case it is clear that, as the referendum date approached, the Leave vote was gaining traction and the proportion of Undecided votes was decreasing.

\section{A random walk model for generating time series with application to poll results}
\label{sec:model}

Stochastic differential equations (SDEs) \cite{MACK11,EVAN13} can provide effective generative models for time series.
In particular, when the SDEs are ``mean reverting'' \cite{HIRS14}, as is the case here, they often possess stationary solutions that fit  various known distributions \cite{COBB81,BIBB05}.
In our application, analysing poll results, the beta distribution \cite{GUPT04} is a natural choice, since it is flexible, designed to model proportions due to its bounded support \cite{KOTZ04}, and is the conjugate prior of the binomial distribution. We also considered the gamma distribution \cite{JOHN94c}, which is a reasonable choice given its relationship to the beta distribution \cite{LEEM08}. However, it only leads to an approximation of the bounded domain and, moreover, it is non-trivial to constrain it to a bounded domain.
Generating beta distribution models using SDEs has applications in other domains, notably in finance \cite{TAUF07}.

\smallskip

A typical {\em stochastic differential equation} (SDE) takes the form
\begin{equation}\label{eq:sde}
{\mathrm d} X_t = \mu(X_t) {\mathrm d} t + \sigma(X_t) {\mathrm d} W_t,
\end{equation}
where $X_t$ is a random variable with $t \ge 0$ a real number denoting time, $\mu(X_t)$ and $\sigma(X_t)$ are known as the {\em drift} and {\em diffusion} functions, respectively, and $W_t$ is a Wiener process (also known as Brownian motion).
Moreover, when
\begin{equation}\label{eq:drift}
\mu(X_t) = \theta \left(m - X_t \right),
\end{equation}
where $\theta$, the {\em rate parameter}, is a positive constant and $m$ is a constant representing the mean of the underlying stochastic diffusion process, the SDE has a stationary solution \cite{COBB81}. In addition, its {\em autocorrelation function} is exponentially decreasing \cite{BIBB05} and takes the form
\begin{equation}\label{eq:autocorr}
\exp(- \theta t).
\end{equation}
\smallskip

It was shown in \cite{COBB81,BIBB05} that, if
\begin{equation}\label{eq:mean}
m = \frac{\alpha}{\alpha + \beta}
\end{equation}
and
\begin{equation}\label{eq:diffusion}
\sigma^2(X_t) = \frac{2 \theta}{\alpha + \beta} \ X_t \left(1 - X_t\right),
\end{equation}
then the marginal distribution of the stationary solution of the SDE is a beta distribution \cite{GUPT04}
with probability density function
\begin{equation}\label{eq:beta}
\frac{\Gamma(\alpha + \beta)}{\Gamma(\alpha) \Gamma(\beta)} x^{\alpha -1} (1-x)^{\beta -1},
\end{equation}
where $\Gamma$ is the gamma function \cite[6.1]{ABRA72}.

\smallskip

Substituting (\ref{eq:drift}) and (\ref{eq:diffusion}) into (\ref{eq:sde}), we obtain the SDE for a diffusion process with a marginal beta distribution in the form
\begin{equation}\label{eq:beta-sde}
{\mathrm d} X_t =  \theta \left(\frac{\alpha}{\alpha + \beta} - X_t \right) {\mathrm d} t +  \sqrt{\frac{2 \theta}{\alpha + \beta} \ X_t \left(1 - X_t\right)} {\mathrm d} W_t.
\end{equation}
\smallskip

We note that several other forms for $m$ and $\sigma^2(X_t)$ also lead to well-known distributions \cite{COBB81,BIBB05}.
Although we maintain that the SDE model we adopt is a natural one in our context, we note that a different model based on Markov chains, which also has a beta distribution as its stationary solution, has been presented in \cite{PACH08}.
In this Markov chain model, at any given time step, the movement in the time series may be up or down with a certain probability. Then the new position, in the interval between $0$ and $1$, is determined according to some density function. Although promising, the results in \cite{PACH08} are not as general as those of the SDE model, and depend on making a choice of parameters that would be difficult to determine from the data.

\medskip

In reality, the continuous SDE model is an approximation of a discrete process described by a stochastic difference equation, where $x_i$
is the discrete analogue of the random variable $X_{t_i}$ at discrete time $t_i$.
Setting $x_0 = X_0$, the dynamics of the discrete process can be described by the difference equation
\begin{equation}\label{eq:em}
\Delta x_{i+1} = \theta \left(\frac{\alpha}{\alpha+\beta} - x_i \right) \Delta t_{i+1} + \sqrt{\frac{2 \theta}{\alpha + \beta} \ x_i \left(1 - x_i\right)} \ \Delta W_{i+1},
\end{equation}
corresponding to (\ref{eq:beta-sde}), where
\begin{equation}\label{eq:dx}
\Delta x_{i+1} = x_{i+1} - x_i,
\end{equation}
\begin{equation}\label{eq:dt}
\Delta t_{i+1} = t_{i+1} - t_i,
\end{equation}
and
\begin{equation}\label{eq:dw}
\Delta W_{i+1} = z_{i+1} \sqrt{\Delta t_{i+1}},
\end{equation}
where $z_{i+1}$ is a normally distributed random variable with mean 0 and variance 1.

\smallskip

Using (\ref{eq:em}) to obtain a computational solution of (\ref{eq:sde}) is known as the {\em Euler-Maruyama method} \cite{SAUE13}, which is a general method for obtaining approximate numerical solutions to SDEs.
We note that this method and various refinements of it are especially useful when analytic solutions do not exist \cite{IACU08}.

\smallskip

In our model of the polls, we assume that the $i$th poll is conducted at time $t_i$, where $t_i = i$.
Thus, in this case, $\Delta t_{i+1}$ in (\ref{eq:dt}) and (\ref{eq:dw}) is taken to be 1.
The proportion of the poll respondents voting for a given outcome, for example Remain, is represented by $x_i$, where $0 \le x_i \le 1$.

\section{Analysis of the Brexit polls considered as a random walk}
\label{sec:uk}

To evaluate the model, we followed a similar approach to that taken in \cite{TAUF07}. We first fit a beta distribution to the marginal distribution of the time series induced by the poll results using the maximum likelihood method to obtain estimates for $\alpha$ and $\beta$. We then used the Jensen-Shannon divergence, defined below, to measure the goodness of fit. Lastly, we fit the autocorrelation function of the time series using least squares nonlinear regression to obtain an estimate for $\theta$. All computations were carried out using the Matlab software package.

\smallskip

The {\em Jensen-Shannon divergence} ({\em JSD}) \cite{ENDR03} is a nonparametric measure of the distance between two distributions $\mathbf{p} = (p_i)$ and $\mathbf{q} = (q_i)$, where $i=1,2,\ldots, n$. The formal definition of the {\em JSD}, which is a symmetric version of the Kullback-Leibler divergence and is based on Shannon's entropy \cite{COVE91}, is given by
\begin{equation}\label{eq:jsd}
JSD(\mathbf{p},\mathbf{q}) = \sqrt{\frac{1}{2 \ln{2}} \ \sum_{i=1}^{n} \left( p_i \ \ln{\frac{2 p_i}{p_i + q_i}} + q_i \ \ln{\frac{2 q_i}{p_i + q_i}} \right)},
\end{equation}
where we use the convention that if $p_i = 0$ or $q_i = 0$, or both, $0 \ln{0}$ and $0 \ln{(0/0)}$ are both defined to be $0$.
(The factor $2 \ln{2}$ is included to normalise the {\em JSD} to be between 0 and 1.)
We observe that the JSD is equal to 0 when $\mathbf{p} = \mathbf{q}$.

\smallskip

\begin{figure}[ht]
\begin{center}
\includegraphics[scale=0.39]{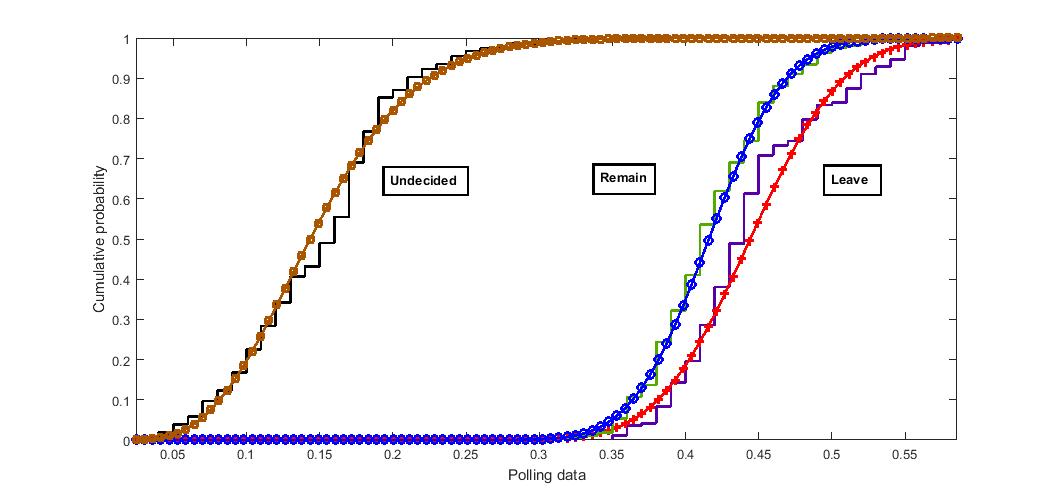}
\caption{\label{figure:beta} Visual presentation of the cumulative fitted beta distributions.}
\end{center}
\end{figure}

In Table~\ref{table:beta} we show the parameters of the beta distribution fitted by the maximum likelihood method, and the JSD between the empirical distribution of the time series of the poll results and the fitted beta distribution. The low JSD values indicate good fits for all three responses.
In Figure~\ref{figure:beta} we show a visual representation, here using {\em cumulative distributions} to highlight the similarities between the empirical and fitted distributions. We note that the fact that the value of the JSD for Leave is somewhat higher is also noticeable from Figure~\ref{figure:beta}.

\begin{table}[ht]
\begin{center}
\begin{tabular}{|l|c|c|c|}\hline
Response    & $\alpha$ & $\beta$  & $JSD$  \\ \hline \hline
Remain      & 59.6781  &  83.6604 & 0.0404 \\ \hline
Leave       & 44.3278  &  55.3813 & 0.0582 \\ \hline
Undecided   &  5.8364  &  33.1904 & 0.0444 \\ \hline
\end{tabular}
\end{center}
\caption{\label{table:beta} Maximum likelihood fitting of the beta distribution to the referendum polls.}
\end{table}

In order to compute the rate parameters $\theta$ of the sample autocorrelation function for the three responses,
we first smoothed the autocorrelation using a moving average filter with a centred sliding window of 5 lags.
We then fitted (\ref{eq:autocorr}) to the smoothed values. The values obtained for $\theta$ are shown in Table~\ref{table:autocorr},
together with the coefficient of determination $R^2$ \cite{MOTU95}, the very high values of which indicate good fits. (We note that using $R^2$ as a goodness-of-fit measure for nonlinear least squares regression is somewhat controversial, although  it has a natural interpretation as the comparison of a given model to the null model \cite{ANDE94}.)

\begin{table}[ht]
\begin{center}
\begin{tabular}{|l|c|c|}\hline
Response    & $\theta$ & $R^2$ \\ \hline \hline
Remain      &  0.9462  & 0.9716 \\ \hline
Leave       &  0.7902  & 0.9393 \\ \hline
Undecided   &  0.9963  & 0.9731 \\ \hline
\end{tabular}
\end{center}
\caption{\label{table:autocorr} Exponential decay autocorrelation parameter of the referendum polls.}
\end{table}

As a demonstration of the predictive power of the model, for each value of $i$, we computed the 95\% confidence interval for the difference between the proportions for the $i$th and $(i+1)$th polls, using (\ref{eq:em});
accordingly, we replaced $z_{i+1}$ in (\ref{eq:dw}) by $\pm 1.96$.
We used the first third of the polls for computing initial values for the parameters $\alpha$ and $\beta$ of the beta distribution, and the rate parameter $\theta$. For the remaining two thirds of the polls and for each response, we next computed the difference between the proportion choosing that response in the poll and the corresponding proportion in the following poll.
We then checked whether this difference was in the computed confidence interval.
After each step we recomputed the values of $\alpha$, $\beta$ and $\theta$ using all the polls up until the current one.
The results are shown in Table~\ref{table:ci-next}, and it can be seen that the predictions for each response were within the appropriate confidence interval over 97\% of the time.

\smallskip

We also computed the difference between the actual result of the referendum and the current poll, to determine whether this difference was in the same confidence interval (this is equivalent to assuming that the following poll was the actual referendum).
It turns out, as can be seen in Table~\ref{table:ci-fin}, that the actual referendum result for Remain was within the predicted confidence intervals in all cases, while this was true for Leave only about 14\% of the time. However, this percentage for Leave increases to 70\% if only the last 20 polls are considered. Thus, even for the supposedly unpredictable referendum result, this is consistent with the adage that the later polls are more informative than the earlier ones.

\begin{table}[ht]
\begin{center}
\begin{tabular}{|l|c|}\hline
Response    &  Proportion in 95\% CI \\ \hline \hline
Remain      &    100\%  \\ \hline
Leave       &  98.23\%  \\ \hline
Undecided   &  97.12\%  \\ \hline
\end{tabular}
\end{center}
\caption{\label{table:ci-next} Percentages of the polls for which the following poll result is within the 95\% confidence interval (CI)
relative to the next step prediction.}
\end{table}

\begin{table}[ht]
\begin{center}
\begin{tabular}{|l|c|}\hline
Response      &  Proportion in 95\% CI \\ \hline \hline
Remain        &  100\%   \\ \hline
Leave         &  14.16\%  \\ \hline
Leave-last 20 &  70\%     \\ \hline
\end{tabular}
\end{center}
\caption{\label{table:ci-fin} Percentages of the polls for which the actual referendum result is within the 95\% confidence interval (CI)
relative to the next step prediction.}
\end{table}

\section{Concluding remarks}
\label{sec:conc}

We have proposed a generative stochastic differential equation model to analyse the time series of poll results; this possesses a stationary solution and the marginal distribution of the time series is a beta distribution. We provided empirical evidence that the model is a good fit to the polls leading up to the Brexit referendum, and also provides good predictive power for the next step prediction task.
We intend investigating other data sets for further validation of the model such as the analysis of polls leading up to a general election.

\newcommand{\etalchar}[1]{$^{#1}$}

\end{document}